%% file: paper.tex
\documentclass[runningheads]{llncs}
\usepackage{graphicx}
\usepackage[ruled,linesnumbered]{algorithm2e}

\usepackage{booktabs} 
\usepackage{amsmath}
\usepackage{xcolor}
\usepackage{url}
\usepackage[titletoc,title]{appendix}

\input{settings}

\begin{document}
\title{Efficient Multiway Hash Join on Reconfigurable Hardware}
%
%
\author{Kunle Olukotun\inst{1} \and Raghu Prabhakar\inst{1} \and
Rekha Singhal\inst{1,2} \and
Jeffrey D.Ullman\inst{1} \and Yaqi Zhang\inst{1}}
\authorrunning{Oluktun et al.}
%
\institute{CS, Stanford University, CA, USA \\
\and
Tata Consultancy Services Research, India\\
\email{\{kunle,raghup17,rekhas2,ullman,yaqiz\}@stanford.com}}
\maketitle              
\begin{abstract}
We propose the algorithms for performing multiway joins using a new type of coarse grain reconfigurable hardware accelerator~-- ``Plasticine''~-- that, compared with other accelerators, emphasizes high compute capability and high on-chip communication bandwidth. Joining three or more relations in a single step, i.e. multiway join, is efficient when the join of any two relations yields too large an intermediate relation. We show at least 200X speedup for a sequence of binary hash joins execution on Plasticine over CPU.  We further show that in some realistic cases, a Plasticine-like accelerator can make 3-way joins more efficient than a cascade of binary hash joins on the same hardware, by a factor of up to 45X.
\end{abstract}
%
%

\input{motivation}

\input{relatedwork}
\input{MJAcceleration}
\input{linear}

\input{cyclic}
\input{Performance}
\input{conclusions}
\appendix
\input{model}

\bibliographystyle{splncs04}
\bibliography{paper}


\end{document}

%% file: settings.tex
\def\BibTeX{{\rm B\kern-.05em{\sc i\kern-.025em b}\kern-.08em
    			 T\kern-.1667em\lower.7ex\hbox{E}\kern-.125emX}}
    			 
\newcommand{\showComments}{yes}

\newcommand{\comment}[2]{
    \ifthenelse{\equal{\showComments}{yes}}{\textcolor{#1}{#2}}{}
}

\newcommand{\ttblue}[1]{\textcolor{blue}{\texttt{#1}}}


%% file: motivation.tex
\section{Motivation} \label{sec:motivation}
Database joins involving more than two relations are at the core of many modern analytics applications.
Examples~\ref{linear-example} and ~\ref{cyclic-example} demonstrate two scenarios that require different types of joins involving three relations.  
\begin{example}
\label{linear-example}
(Linear 3-way join)
Consider queries involving the Facebook ``friends'' relation $F$.
One possible query asks for a count of the ``friends of friends of friends'' for each of the Facebook subscribers, perhaps to find people with a lot of influence over others.  There are approximately two billion Facebook users, each with an average of 300 friends, so $F$ has approximately $6\times 10^{11}$ tuples.  Joining $F$ with itself will result in a relation with approximately $1.8\times 10^{14}$ tuples.\footnote{Technically, there will be duplicates, because if $x$ is a friend of a friend of $y$, then there will usually be more than one friend that is common to $x$ and $y$. But eliminating duplicates is itself an expensive operation. We assume duplicates are not eliminated.
  }  However, the output relation only involves 2 billion tuples, or 1/90000th as much data.\footnote{There is a technical difficulty with answering this query using parallel processing: we must take the union of large, overlapping sets, each produced at one processor.  We cannot avoid this union if we are to get an exact count of the number of friends of friends of friends.  However, we can get an asymptotically correct approximation to the size of the union using a very small amount of data to represent each set.  One method to do so is the algorithm of Flajolet-Martin~\cite{FM} \cite{MMDS}.}  Thus, a three-way join of three copies of $F$ might be more efficient, if we can limit the cost of the input data replication as we execute the three-way join.
\end{example}

\begin{example}
\label{cyclic-example}
(Cyclic 3-way join)
Consider the problem of finding triangles in relation $F$.  That is, we are looking for triples of people who are mutual friends.  The density of triangles in a community might be used to estimate its maturity or its cohesiveness.  There will be many fewer triangles than there are tuples in the join of $F$ with itself, so the output relation will be much smaller than the intermediate binary joins.
\end{example}

Afrati and Ullman~\cite{afrati} showed that in some cases, a multiway join can be more efficient than a cascade of
 binary joins, when implemented using MapReduce. But multiway joins are superior only when the intermediate products (joins of any two relations) are large compared to the required replication of the input data at parallel workers, and the output is relatively small; that is the case in each of the Examples~\ref{linear-example} and ~\ref{cyclic-example}.  The limitation on the efficiency of any parallel algorithm for multiway joins is the degree to which data must be replicated at different processors and  the available computing capacity. The performance benefits of multiway joins over cascaded binary joins could be perceived on hardware architectures facilitating cheap data replication.

Spatially reconfigurable architectures~\cite{spatial}, such as Coarse-grained reconfigurable architecture (CGRA), have gained traction in recent years as high-throughput, low-latency, and energy-efficient accelerators.  
With static configuration and explicitly managed scratchpads, reconfigurable accelerators dramatically reduce energy and performance
overhead introduced by dynamic instruction scheduling and cache hierarchy in CPUs and GPUs.
In contrast
to field-programmable gate arrays (FPGAs), CGRAs are reconfigurable at word or higher-level as opposed to bit-level.
The decrease in flexibility in CGRA reduces routing overhead and improves clock frequency, compute density, and energy-efficiency
compared to FPGAs.

\begin{figure}
  \centering
  \includegraphics[width=0.8\columnwidth]{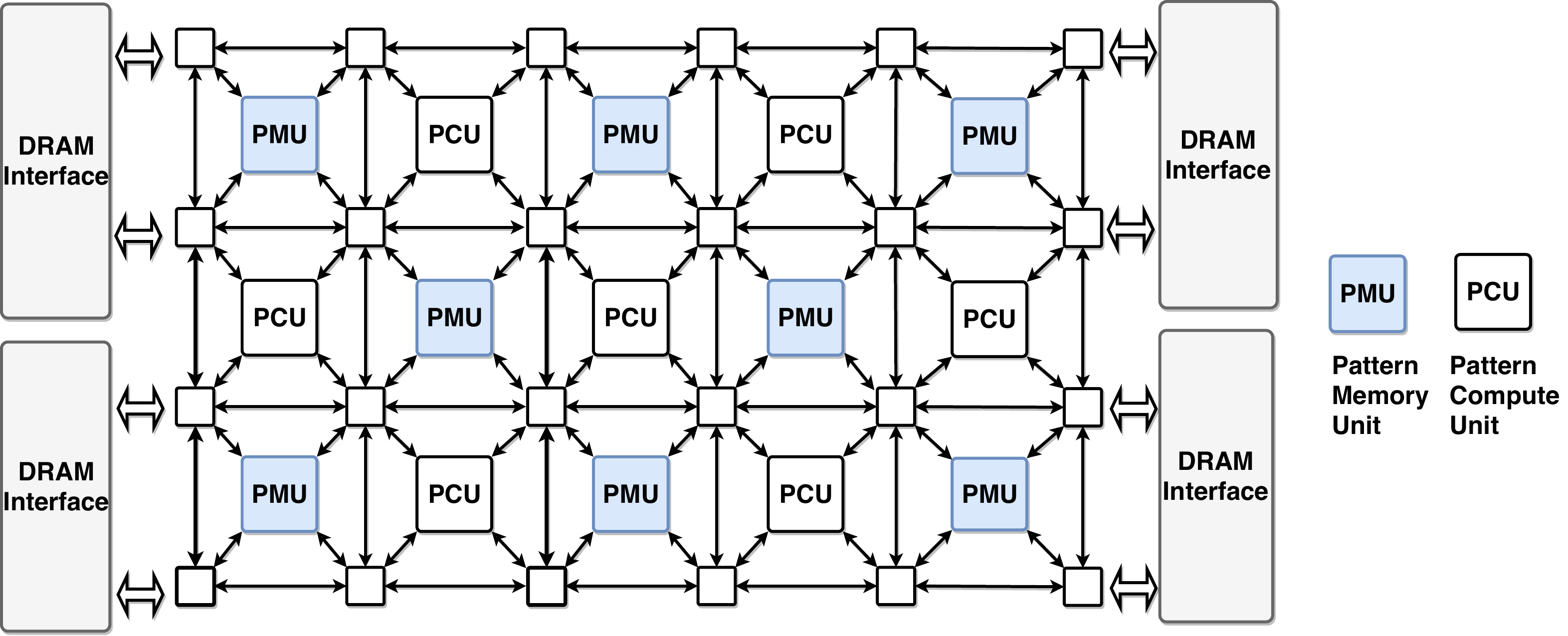}
\caption{Plasticine-like coarse grain reconfigurable hardware accelerator.}
\label{fig:plasticine}
\vspace{-20pt}
\end{figure}

Plasticine~\cite{plasticine} is a recently proposed tile-based CGRA accelerator.  As shown in Fig~\ref{fig:plasticine}, Plasticine has a checkerboard layout of compute and memory units connected with high bandwidth on-chip network.  Plasticine-like architectures offer several advantages to enable efficient multiway join acceleration.  First, it has peak 12.3 FLOPS throughput designed for compute-intensive applications, like multiway join.  Second, the high-bandwidth static network can efficiently broadcast data to multiple destinations, which makes replication very efficient.
\subsection{Contributions}
In this paper, we study algorithms to efficiently perform multiway joins on Plasticine-like accelerator. We show an advantage of such accelerators over CPU-based implementation on a sequence of binary hash joins, and additional performance improvement with 3-way joins over cascaded binary joins.  Although we describe the algorithms with Plasticine as a potential target, the algorithms can also be mapped onto other reconfigurable hardware like FPGAs by overlaying Plasticine structure on top of the substrate architecture.
The contributions of the paper are summarized below.
\begin{itemize}
        \item Algorithms and efficient implementations for both linear and cyclic 3-way join operations for Plasticine-like accelerators.  These algorithms are significantly different from the algorithms of \cite{afrati} for the MapReduce implementation of the same joins.
       
    \item Analysis of the cost of running these algorithms in terms of the number of tuples that are read onto an accelerator chip.
    \item Performance comparison of a sequence of binary hash-join implementation on a Plasticine-like accelerator to state-of-the-art CPU hash-join on Postgres~\cite{stonebraker1986design}.
    \item Evaluation of the 3-way join algorithms compared to the cascaded binary hash-join implementation on the same accelerator.  
\end{itemize}
\subsection{Simplifying Assumptions}
In our analyses, we shall assume a uniform distribution of join-key values.  This assumption is unrealistic because there is typically {\em skew}, where some values appear more frequently than others.  Small amounts of skew can be handled by leaving some components of the accelerator chip to handle ``overflow'' of other components.  However, large amounts of skew require a modification to the algorithms along the lines of~\cite{afrati-skew}, which we do not cover in detail due to space limitation. 

The rest of this paper is organized as follows: Section~\ref{sec:related} presents some background and related work.   Sections~\ref{sec:mjacc} discuss the challenges for multiway join algorithm implementation on Plasticine-like accelerator. Sections~\ref{sec:linear} and \ref{sec:cyclic} present our algorithms for linear and cyclic multiway joins respectively. Section~\ref{sec:perf} compare the performance results of a sequence of binary hash joins on Plasticine-like accelerator and CPU. Further, we also compare the performance of the accelerated multiway join algorithms to an accelerated sequence of binary join approach on Plasticine-like accelerator. Finally the paper concludes with the future work in Section~\ref{sec:conclusions}.

%% file: relatedwork.tex
\section{Background And Related Work} \label{sec:related}
This section provides a brief background and reviews relevant related work on multiway join algorithms, hash-join acceleration, and spatially reconfigurable architectures.

\subsection{Multiway joins}
Efficient join algorithms are usually based on hashing~\cite{hash-sort}. Parallelism can be exploited by the parallel processing of a tree of several binary joins~\cite{multiplan}, an approach that is unsuitable for joins generating large intermediate relations, as is the case for our two introductory examples. The focus of such approaches has been to find optimal plans for parallel execution of binary joins. Henderson et al.~\cite{MJanalysis} presented a performance comparison of different types of multiway-join structures to two-way (binary) join algorithm. 

A leapfrog approach~\cite{leapfrog} has been used to join multiple relations simultaneously by parallel scanning of the relations that are sorted on the join key. Aberger et al.~\cite{emptyheaded} have accelerated the performance of leapfrog triejoin using SIMD set intersections on CPU-based systems. The algorithm is sequential on the number of join keys and requires the relations to be preprocessed into trie data structures.
\subsection{Hash-Join Acceleration}
A hash-join algorithm on large relations involves three key operations - partitioning of relations, hashing of the smaller relation into a memory (build phase) followed by the probing of the second relation in the memory. Kara et al.~\cite{partition} present an efficient algorithm for partitioning relations using FPGA-based accelerator.
Onur et al.~\cite{walker} use on-chip accelerator for hash index lookup (probing) to process multiple keys in parallel on a set of programmable 'walker' units for hashing. 
Robert et al.~\cite{fpgahashjoin,FPGAmultithread} use FPGA for parallelizing hashing and collision resolution in the building phase. 
Huang et al.~\cite{mjaccl} have explored the use of open coherent accelerator processor interface (OpenCAPI)-attached FPGA to accelerate 3-way multiway joins where the intermediate join of two relations is pipelined with a partition phase and join with the third relation. 
\subsection{Spatially Reconfigurable Architectures}
Spatially reconfigurable architectures are composed of reconfigurable compute and memory blocks that are connected to each other using a programmable interconnect. Such architectures are a promising compute substrate to perform hardware acceleration, as they avoid the overheads in conventional processor pipelines, while retaining the flexibility. Recent work has shown that some spatially reconfigurable architectures achieve superior performance and energy efficiency benefits over fine-grained alternatives such as FPGAs and conventional CPUs~\cite{plasticine}.

Several spatially reconfigurable architectures have been proposed in the past for various domains. Architectures such as Dyser~\cite{dyser} and Garp~\cite{garp} are tightly coupled with a general purpose CPU. Others such as Piperench~\cite{piperench}, Tartan~\cite{tartan}, and Plasticine~\cite{plasticine} are more hierarchical with coarser-grained building blocks. Plasticine-like accelerator is not limited to databases alone but can efficiently accelerate multiway joins. Q100~\cite{Q100} and Linqits~\cite{linqits} are accelerators specific to databases. 

%% file: MJAcceleration.tex
\section{Accelerating Multiway Joins} \label{sec:mjacc}
We present algorithms for accelerating both linear ($R(AB)\bowtie S(BC)\bowtie T(CD)$) and cyclic ($R(AB)\bowtie S(BC)\bowtie T(CA)$) multiway joins on a Plasticine-like accelerator using hashing. There may be other attributes of relations $R$, $S$, and $T$. These may be
 assumed to be carried along as we join tuples, but do not affect the algorithms.   Also, $A$, $B$, $C$, and $D$ can each represent several columns of the relations and by symmetry, assume that $|R|\le|T|$.

A naive approach to map the Afrati et al.~\cite{afrati} algorithm on Plasticine-like architecture will be bottlenecked by DRAM bandwidth and limited by the size of on-chip memory. The proposed multiway hash-join algorithms exploit the pipeline and parallelism benefits in a Plasticine-like architecture to improve the performance while eliminating the limitations mentioned above.

We partition one or more relations using hash functions, one for each of the columns used for joining, such that the size of potentially matching partitions of the three relations is less than or equal to the size of on-chip memory. The loading of a partition of a relation from DRAM to on-chip memory is pipelined with the processing of the previously loaded partition(s) on the accelerator. Further, to squeeze more processing within the given on-chip memory budget, at least one of the relations is streamed, unlike batch processing in Afrati et al.\cite{afrati}.
\input{Notation.tex}

%% file: Notation.tex
\subsection{Notations}
In what follows, we use $|R|$ to represent the number of records of a relation R. A relation $R(AB)$'s tuple is represented as $r(a,b)$ and the column $B$'s values is accessed as $r.b$.
We use the name of hash functions--$h$, $g$, $f$, $G$, and $H$ (or $h_{bkt}$, $g_{bkt}$, $f_{bkt}$, $G_{bkt}$, and $H_{bkt}$) in certain equations to stand for the number of buckets produced by those functions.
$U$ is the number of distributed memory and compute units, and we assume there is an equal number of each.  $M$ is the total on-chip memory capacity.

%% file: linear.tex
\section{Linear 3-Way Join}\label{sec:linear}
\begin{figure}[htbp]
\includegraphics[width=0.98\linewidth]{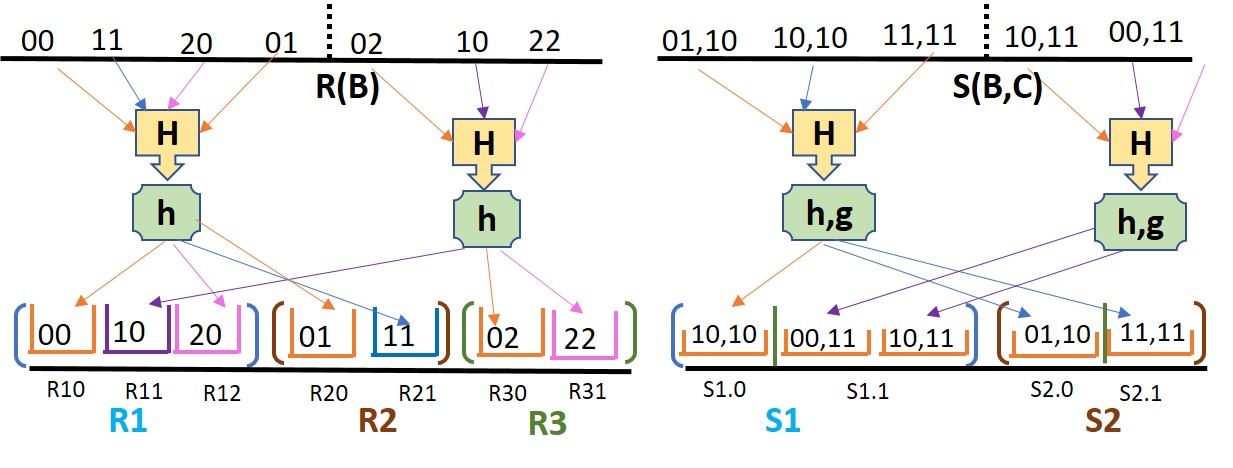}
\caption{Partitioning of Relation $R$ and $S$. Relation $R$ is partitioned using radix hashing on the first digit, $H(B)$, to create subpartitions $R_1$, $R_2$, and $R_3$. Each $R_i$ is further partitioned using radix hashing, $h(B)$, on the second digit of $B$. $S$ is partitioned using radix hashing similar to $R$, on both the $B$ and $C$ columns.}
\label{fig:sys}
\end{figure}

For the linear, three-way join $R(AB) \bowtie S(BC) \bowtie T(CD)$, we partition the relations at two levels in a particular way, using hash functions as shown in Fig~\ref{fig:sys}. The relations are partitioned using robust hash functions~\cite{partition-study} on the columns involved in the join, which, given our no-skew assumption, assures uniform sizes for all partitions.
We can first configure the accelerator to perform the needed partitioning.  Since all hash-join algorithms require a similar partitioning, we shall not go into details regarding the implementation of this step.
\begin{algorithm}[!htp]
 \KwData{Relations R(A,B), S(B,C) and T(C, D). Memory grid, MemGrid[], on accelerator. Column B values hashed using $H()$ and $h()$, and Column C hashed using $g()$. $\#Rpart$ denotes the number of partitions of relation $R$}
 \KwResult{Tuples from R, S and T joined on common values of B and C.}
 $T_i \leftarrow$ Partition T(C,D) using hash function g(C) $[\#Tpart]$\;
 $S_{ij} \leftarrow$ Partition S(B,C) using hash function H(B) and g(C)  (Sij partitions are ordered first on H(B) and then on g(C) within each Si partition, $[\#S_{i}part]$) \;
 $R_i \leftarrow$ Partition R(A,B)using hash function H(B) $[\#Rpart]$\;
\For{Each partition $R_{i=H(B)}$ of R till $\#Rpart$}{\
 \For{All records of $R_i$}{\
  $h_{b} \leftarrow h(r_i.b)$\;
  $MemGrid[h_{b}] \leftarrow r_i(*,b)$\;
  } 
 \For{Each partition $S_{i=H(B)}$ of S till \#Spart}{\
     \For{Each partition $S_{ij=g(C)}$  till \#$S_{i}$part}{\
       \For{All records of $S_{ij}$}{
        $h_{b} \leftarrow h(s_{ij}.b)$\;
        $MemGrid[h_{b}] \leftarrow s_{ij}(b,c)$\;
        } 
       $MemGrid[*]  \leftarrow t_j(c,*)$ [broadcast or send to all Memory units where $S_{ij}$ was sent]\;
       Join tuple from $R_i$, $S_{ij}$ and $T_j$\;
       Discard tuples from $S_{ij}$ and $T_j$\;
     } 
  }
  Discard tuples of $R_i$
} 
  \caption{Pseudo-code for $R(AB) \bowtie S(BC) \bowtie T(CD)$}
 \label{algo:linearjoin1}
\end{algorithm}

The relations $R$ and $T$ are similar, each having one join column, while relation $S$ has two columns to join with relations $R$ and $T$. The relative sizes of the three relations affect our choice of algorithm.  The largest relation should be streamed to the accelerator to optimize the on-chip memory budget. When $S$ is largest, relations $R$ and $T$ must either be small enough to fit on the on-chip memory (discussed in detail as a ``star'' 3-way join in Section~\ref{sec:perf}) or they should be partitioned, based on the values of attributes $B$ or $C$, respectively, each of them having $L$ sub-partitions. Then each pair of sub-partitions is loaded on to the accelerator iteratively and matched with the corresponding one of the $L^2$ partition of the streamed relation $S$.  In the case of larger $R$ and $T$ relations, one of them is streamed and the other one is partitioned as discussed in detail below.
\subsection{Joining Relations on Plasticine-like Accelerator}
Consider the case where $S$ is no larger than $R$ or $T$. 
For the first level partitioning of the relations $R$ and $S$ on attribute $B$,  we choose a number of partitions for the hash function $H(B)$ so that a single partition of $R$ (that is, the set of tuples of $R$ whose $B$-value hashes to that partition) will fit comfortably in one pattern memory unit (PMU) of the Plasticine. 
The second level of partitioning serves two purposes and involves two hash functions.  First, we use hash function $h(B)$ to divide a single partition of $R$ and $S$ into $U$ buckets each, one bucket per PMU.
We use hash function $g(C)$ to divide $C$ into a very large number of buckets.  Each partition of $S$ is further partitioned into sub-partitions that correspond to a single value of $g(C)$. Each $g(C)$ bucket of $S$'s partition may be organized by increasing values of $h(B)$ as shown in Fig~\ref{fig:sys}. Likewise, the entire relation $T$ is divided into buckets based on the value of $g(C)$.

We shall describe what happens when we join a single partition of $R$, that is, the set of tuples of $R$ whose $B$-values have a fixed value $H(B)=i$, with the corresponding partition of $S$ (the set of tuples of $S$ whose $B$-values also have $H(B)=i$.  Call these partitions $R_i$ and $S_i$, respectively.
\begin{enumerate}
    \item Bring the entire partition of $R$ onto the chip, storing each tuple $r(a,b)$ in the PMU for $h(b)$.
    \item For each bucket of $g(C)$, bring each tuple $s(b,c)$ from that bucket from $S_i$ onto the chip and store it in the PMU for $h(b)$.
    \item Once the bucket from $S_i$ has been read onto the chip, read the corresponding bucket of $T$~-- $t(c,d)$ with the same hash value $g(C)$~-- onto the chip. Since tuple $t(c,d)$ can join with tuple $r(a,b)$ and $s(b,c)$ having any value of $B$, we must route each $t(c,d)$ tuple to every PMU.
    \item Once the buckets with a given value $g(C)$ have arrived, PCUs joins the three tiny relations at each PMU using optimized cascaded binary joins. Recall we assume the result of this join is small because some aggregation of the result is done, as discussed in Example~\ref{linear-example}.  Thus, the amount of memory needed to compute the join at a single memory is small.\footnote{For just one example, if $R$, $S$, and $T$ are each the friends relation $F$, and we are using the Flajolet-Martin algorithm to estimate the number of friends of friends of friends for each individual $A$ in the relation $R$, then the amount of data that needs to be maintained in memory would be on the order of 100 bytes for each tuple in the partition $R_i$, and thus would not be more than proportional to the size of the data that was read into the memory from outside.  In fact, although we do not want to get into the details of the Flajolet-Martin algorithm~\cite{MMDS}, if we are willing to assume that everyone has at least some small number of friends of friends of friends, e.g., at least 256, then we can reduce the needed space per tuple to almost nothing.}
\end{enumerate}
The formal representation of the algorithm is presented in Algorithm~\ref{algo:linearjoin1}.
\subsection{Analysis of the Linear 3-way Join}
Each tuple of $R$ and $S$ is read onto an accelerator chip exactly once.  However, tuples of $T$ are read many times~-- once for each partition of $R$.  The number of partitions produced by the hash function $H(B)$ is such that one partition of $R$ fits onto the entire on-chip memory with capacity $M$. Thus, the number of partitions into which $R$ is partitioned is $\frac{|R|}{M}$. Therefore, the number of reads for tuples of $T$ is $\frac{|R||T|}{M}$.  This function is symmetric in $R$ and $T$, so it seems not to matter whether $R$ is the smaller or larger of the two relations.  However, we also have to read $R$ once, so we would prefer that $R$ be the smaller of $R$ and $T$. That is, the total number of tuples read is $|R|+|S|+\frac{|R||T|}{M}$.

Thus, the number of tuples read onto the chip is greater than the sizes of the three relations being joined.  However, using a cascade of two-way joins may also involve an intermediate relation whose size is much bigger than the sizes of the input relations.  Thus, while we cannot be certain that the three-way join is more efficient than the conventional pair of two-way joins, it is at least possible that the algorithm proposed will be more efficient.

\begin{example}
\label{ex:friends}
Consider again the problem of getting an approximate count of the friends of friends of friends of each Facebook user, as was introduced in Example~\ref{linear-example}. We estimated the number of tuples in the friends relation $F$ as $6\times10^{11}$.  This value is thus the sizes of each of $R$, $S$, and $T$.  If we take the three-way join, then the number of tuples read onto an accelerator chip is $6\times10^{11} + 6\times10^{11} +3.6\times10^{23}/M$.  In comparison, if we use two two-way joins, then we need to output first the join of $F$ with itself, which involves producing about $1.8\times10^{14}$ tuples, and then reading these tuples back in again when we join their relation with the third relation.  The three-way join will involve reading fewer tuples if $6\times10^{11} + 6\times10^{11} +3.6\times10^{23}/M < 3.6\times10^{14}$.  That relationship will hold if $M > 1.003\times 10^9$.  That number is far more than can be expected on a single chip with today's technologies, even
  assuming that a tuple is only eight bytes (two 4-byte integers representing a pair of user ID's).  However, for somewhat smaller databases, e.g., the 300 million Twitter users and their followers, the on-chip memory requirements are feasible, in that case, the chip needs to hold approximately 150 million tuples.\footnote{In fact, as a general rule, we can observe that the minimum memory size $M$ needed for any social-network graph is very close to half the number of nodes in the graph, regardless of the average degree of the graph (number of friends per user) and size of the relation.}
\end{example}

%% file: cyclic.tex
\section{Cyclic 3-Way Join}
\label{sec:cyclic}
Consider the cyclic three-way join $R(AB)\bowtie S(BC)\bowtie T(CA)$.  
The cyclic join is symmetric in all three relations.  We shall therefore assume that $R$ is the smallest of the three, for reasons we shall see shortly.
Similar to the linear three-way join, we shall partition $R$ such that it's one partition fits conveniently into on-chip memory.  However, in this case, since both $A$ and $B$ are shared by other relations, we will partition $R$ using hash functions $H(A)$ and $G(B)$ into $H$, and $G$ buckets, respectively.  The correct values of $H$ and $G$ are to be determined by considering the relative sizes of the three relations.  However, we do know that $\frac{|R|}{HG} = M$. 
\begin{figure}
\includegraphics[width=1\columnwidth,trim={0 7cm 0 5cm},clip]{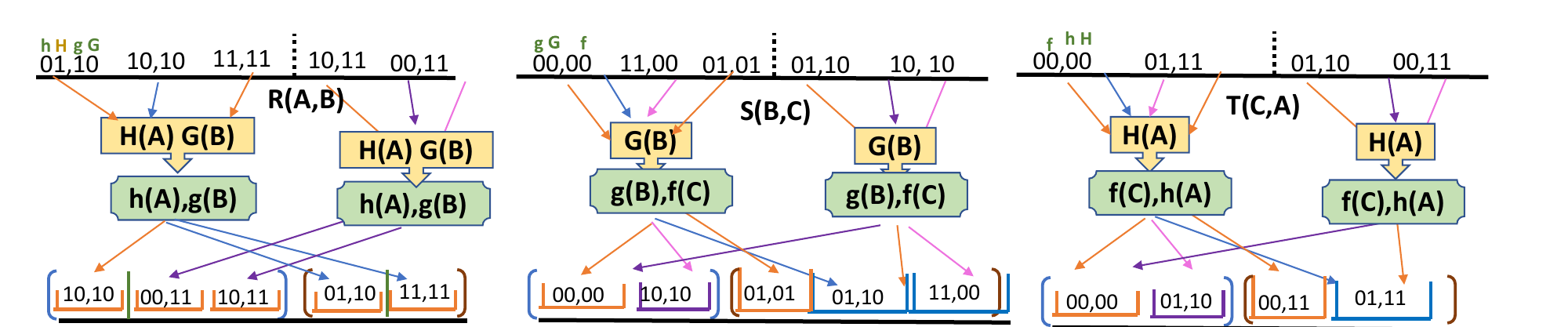}
  \caption{Partitioning of Relation $R$, $S$ and $T$. Relation $R$ is partitioned using radix hashing on the first digit of column A and B using $H(A), G(B)$ respectively. Each $R_i$ is further partitioned using radix hashing, $h(A),g(B)$, on the second digit of $A, B$. Similarly, $S$ and $T$ are partitioned using radix hashing on $B$ and $A$ columns respectively. Column $C$ is hashed using $f(C)$.}
  \label{fig:cyclic}
  \end{figure}
In addition to partitioning $R$ into $HG$ pieces, each of size $M$, we use $H(A)$ to partition $T$ into $H$ pieces, each of size $\frac{|T|}{H}$, and we use $G(B)$ to partition $S$ into $G$ pieces, each of size $\frac{|S|}{G}$. The partitioning scheme is depicted in Fig~\ref{fig:cyclic}. 

As before, we are assuming that there is no significant skew in the distribution of values in any column, and we also are assuming a sufficient number of different values that hashing will divide the relations approximately evenly.  In what follows, we shall only describe the join of a single partition from each of $R$, $S$, and $T$.  These three partitions are determined by buckets of $H$ and $G$.  That is, for a fixed value of $H(A)=i$ and a fixed value of $G(B)=j$, we join those tuples $r(a,b)$ of $R$ such that $H(a)=i$ and $G(b)=j$ with the tuples $s(b,c)$ of $S$ such that $G(b)=j$ and the tuples $t(c,a)$ of $T$ such that $H(a)=i$.  In what follows, we shall refer to these partitions as $R'$, $S'$, and $T'$, respectively.  Each set of three partitions is handled the same way, either sequentially on one accelerator chip or in parallel on more than one such chip.
\subsection{Joining Relations on Plasticine-like Accelerator}
Now, let us focus on joining $R'$, $S'$, and $T'$.  Assuming the chip has $U$ memories arranged in a square $\sqrt{U}$ on a side, we shall use lower-level hash functions $h(A)$, $g(B)$, and $f(C)$.  Hash functions $h$ and $g$ each map to $\sqrt{U}$ buckets, while $f$ maps to a very large number of buckets~-- a sufficient number of buckets so that $S'$ and $T'$ can be partitioned on the basis of their $C$-values into pieces that are sufficiently small that we can neglect the memory space needed to store one piece from one of these two relations.

Begin the join by bringing onto the chip all the tuples $r'(a,b)$ of $R'$.  Each of these tuples is routed to only one of the $U$ PMUs~-- the PMU in row $h(a)$ and column $g(b)$. 
Then we bring onto the chip each of the tuples $s'(b,c)$ of $S'$ that have $f(c)=k$.  These tuples are each stored in every PMU in the column $g(b)$.  Thus, this tuple will meet at one of these memories, all the tuples of $R'$ that share the same hash value $g(B)$.  Finally, we pipe in the tuples $t'(c,a)$ of $T'$ that have $f(c)=k$.  Each of these tuples is read into each of the memories in row $h(a)$, where it is joined with the possibly matching tuples $r'(a,b)$ and $s'(b,c)$.  Any matches are sent to the output of the chip.

\subsection{Analysis of Cyclic Three-Way Join}
Notice first that every top-level partition of $R$ is read onto the chip only once.  However, a top-level partition of $S$ is read onto chip $H$ times, once for each bucket of $H(A)$.  Also, every top-level partition of $T$ is read $G$ times, once for each bucket of $G(B)$.  The total number of tuples read onto an accelerator chip is thus $|R| + H|S| + G|T|$.
Recall also that $GH=\frac{|R|}{M}$, so previous function can be expressed as
$|R| + H|S| + \frac{|R||T|}{MH}$.
To minimize this function, set its derivative with respect to $H$ to 0, which gives us
$H = \sqrt{\frac{|R||T|}{M|S|}}$.  For this value of $H$, the cost function becomes 
$|R| + 2\sqrt{\frac{|R||S||T|}{M}}$.
Notice that the second term is independent of the relative sizes of the three relations, but the first term, $|R|$, tells us that the total number of tuples read is minimized when we pick $R$ to be the smallest of the three relations.

\begin{example}
  Suppose each of the three relations is the Facebook friends relation $F$; that is, $|R|=|S|=|T|=6\times10^{11}$.  Then the total number of tuples read onto the chip is $6\times10^{11}(1+\sqrt{6\times10^{11}/M})$.  If we assume as in Example~\ref{ex:friends} that the binary join of $F$ with itself has about $0.8\times10^{14}$ tuples, we can conclude that the total number of tuples read by the three-way join of $F$ with itself is less than the number of tuples produced in the intermediate product of two copies of a cascade of two-way joins as long as $6\times10^{11}(1+\sqrt{\frac{6\times10^{11}}{U}})<1.8\times10^{14}$.
This condition is satisfied for $M$ as small as seven million tuples.
\end{example}

%% file: Performance.tex
\section{Performance Evaluation} \label{sec:perf}
In this section, we evaluate the algorithms proposed in the Sections~\ref{sec:linear}, on Plasticine-like accelerator using a performance model.
First, we show the advantage of accelerating a sequence of binary join operators by comparing its execution time on Postgres database on CPU to our simulation on the accelerator. Next, we show additional
performance improvement of 3-way join (an instance of multiway join) over a cascade of two binary hash joins on the acccelerator.

We consider two categories of multiway joins in this evaluation: self-join\footnote{Self 3-way join is joining of a relation with two instances of itself e.g. Friends of friends.} 
of a big relation of size $N$, where N does not fit on-chip; and star-join\footnote{Star 3-way join is joining of a large fact relation with two small dimension relations e.g. TPCH~\cite{tpch} benchmark having join of {\em lineitem} fact relation with {\em order} and {\em supplier} dimension relations.} 
of two small relations ($R$ and $T$) each of size $K$ with a large relation, $S$, of size $N$, 
where $N >> K$ and $2K <= M$. 
The self join algorithm described in Section~\ref{sec:linear} is a generic
algorithm for any linear join, whereas the algorithm used for star join is a variant of the generic algorithm
that specialize for better locality when the dimension relations fit on the on-chip memory.

For a given set of relations, we observe that the proposed algorithms execution time on the accelerator is sensitive to the number of buckets and DRAM bandwidth. We first evaluate the selection of hyperparameters of the algorithms, i.e. bucket size
for the cascaded binary and 3-way joins. With best bucket sizes, 
we compare the performance advantage of 3-way join over a cascade of binary joins for different selectivity of join columns and
DRAM-bandwidths. For all experiments, we do not materialize the final output of the join in memory (refer Example~\ref{linear-example}). Instead,
we assume the final results will be aggregated on the fly. 
Therefore, in our study, we only materialize the intermediate result of the first
binary join, and the final output is immediately aggregated (e.g. perform count operation on the number of friends of friends relation).
\subsection{Target Systems}\label{sec:target}
The CPU system, used for performance evaluation of cascaded binary join, is Intel Xeon Processor E5-2697 v2 server with 30M shared cache and 251GB DDR3 RAM. Although the platform is multi-processor, the Postgres implementation
is single threaded. Nonetheless, we do make sure no other application is contending for 
DRAM bandwidth.
For performance evalutaion on hardware accelerator, we use performance model for the Plasticine-like architecture. It  has DDR3 DRAM technology with 49GB/s read and write bandwidth , Number of PMUs(PCUs), $U=64$ and a peak of
12.3 TFLOPS compute throughput with 16MB on-chip scratchpad. 
\subsection{Accelerator's Performance Model}
The performance model is built by simulating the logic of the proposed algorithm on the hardware specification of the accelerator given in Section~\ref{sec:target}.
We observed that the performance advantage of the proposed 3-way join over cascaded binary join depends on the number of records in the joining relations and the selectivity of the join column - lower selectivity (i.e. higher duplicates) favors multiway join. The performance model needs two inputs for simulation - the number of records of $R$, $S$ and $T$ and the maximum distinct values over all joining columns  (represented as $d$).

The performance model accounts for how an application is spatially parallelized and data is streamed across compute and memory units of the accelerator. The model does considers DRAM-contention while loading multiple data streams concurrently on the chip.
For higher DRAM bandwidth utilization and to hide the DRAM latency, we overlap execution of the algorithm with prefetching of the data. This requires to split the on-chip memory into two buffers (double buffering) to store
both the current and prefetched data. 
The performance model uses only half of the on-chip memory to include this optimization.

For cascaded binary join, once the intermediate
result does not fit in DRAM, the performance model simulates the flushing of the intermediate data to the underlying persistent storage with much lower bandwidth (around 700MB/s from the latest SSD technology).
Appendix~\ref{sec:model} explains the performance model in detail.
\subsection{Performance Analysis of Cascaded Binary Join}\label{sec:selfjoin2}
A cascaded binary-join is a sequence of two binary joins- the first join is $R(AB) \bowtie S(BC)$ which outputs intermediate relation $I(ABC)$ and second join is $I(ABC) \bowtie T(CD)$. For uniform distribution, the intermediate size for a cascaded binary join is $|I|= |R\bowtie S|\leq\frac{|R||S|}{d}$ \cite{swami1994estimation}.


Both the joins are executed on the accelerator similar to the 3-way join discussed in Section~\ref{sec:linear}.
The first join $R(AB) \bowtie S(BC)$  involves loading and matching of partitions of $R$ and $S$ using $H(B),h(B)$ on the chip. The intermediate relation $I$ is copied from  on-chip to the DRAM. The second join  $I(A,B,C)\bowtie T(C,D)$ is identical except the output results are no longer materialized in DRAM. 
For the second join, we also load partitions of relation $T$ on-chip while streaming 
previous join intermediate result, since $|R\bowtie S| >> |T|$.
The bucket sizes of the second level hash functions for both the joins are fixed to the number of PMUs, i.e. $h=g=U$.

Fig~\ref{fig:perf} (a) shows the breakup of the execution time of a cascaded binary self join of three relations with a varying number of buckets i.e. $H_{bkt}$.
The orange region shows time spent in partitioning the relations for both the joins, which is dominated
by the second join due to large size of the intermediate relation. Clearly, the
first join is bounded by DRAM-bandwidth, varying $H_{bkt}$ has no impact on the performance. Fig~\ref{fig:perf} (b) shows variation of 
the execution time of the second join varying $G_{bkt}$. The second join is compute-bound at small $G_{bkt}$
, as the total amount of data loaded is $|R \bowtie S| + |T|$, whereas the total comparison is $\frac{|R \bowtie S||T|}{d}$.
\subsubsection{Performance comparison of Cascaded Binary Join}\label{sec:selfjoin2cpu}
We compare the performance of cascaded binary join on CPU to that on the accelerator using configuration given in Section~\ref{sec:target}. For CPU-based implementation, we follow a \texttt{COUNT} aggregation
immediately after the cascaded binary joins, which prevents the final output to be materialized in memory.
We configure the accelerator with the same DRAM capacity as our baseline CPU.

\begin{figure}
\includegraphics[width=1\columnwidth]{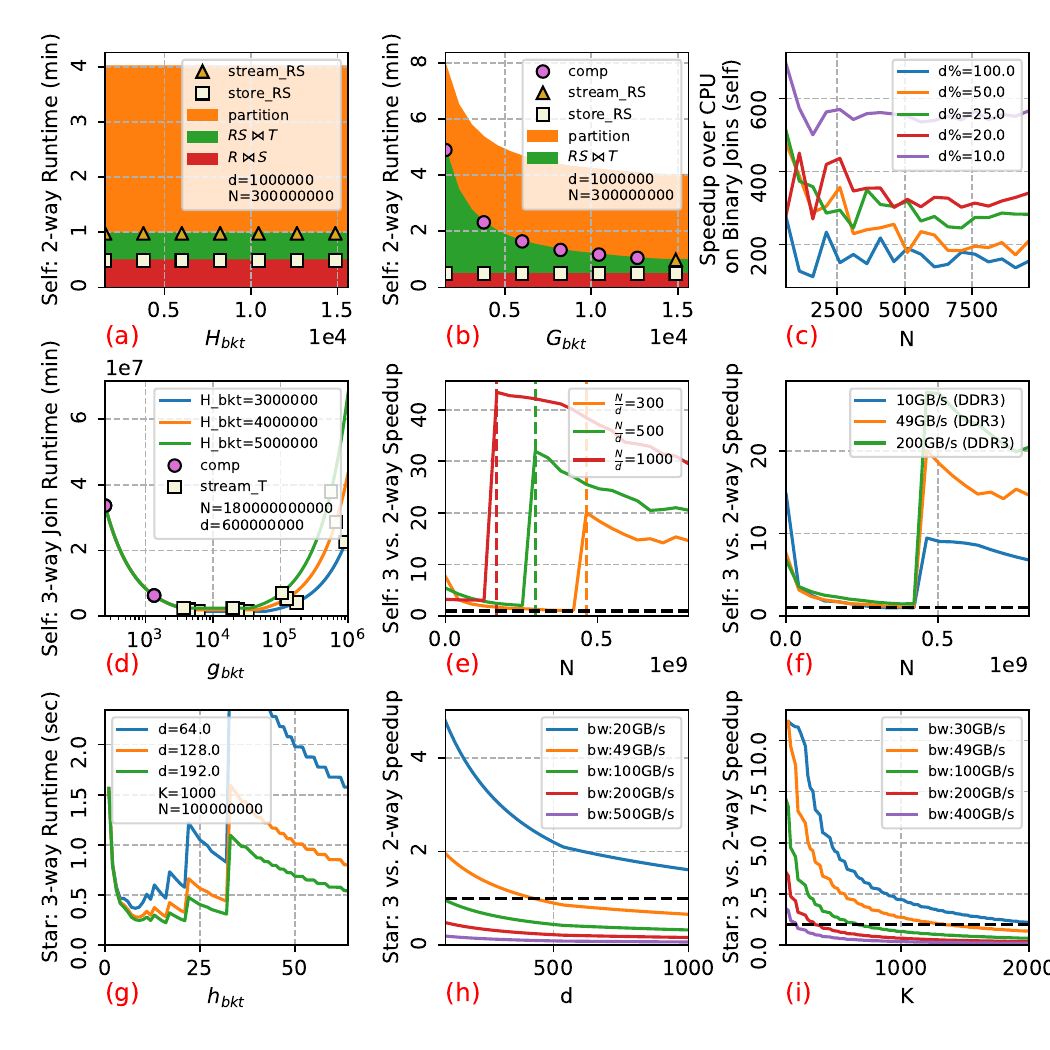}
  \caption{Performance Evaluation of 3-way join vs. cascaded binary joins. 
  (a,b) 2-way
  self linear join  execution time with breakup. Red, green, and yellow region indicate execution time for the first join,
  the second join, and partitioning time for both.
  Marker indicates performance bottleneck in computation (comp), 
  streaming in $R \bowtie S$ relation in second join (stream\_RS), or storing $R \bowtie S$ in first join (store\_RS).
  (c) Speedup on Plasticine over CPU for cascaded binary self joins.
  (d) 3-way linear self join performance. 
  Marker indicates bottleneck of performance in computation (comp) or streaming in T relation (stream\_T). 
  (e) Speedup of 3 vs. binary join on linear self join with DDR3 and SSD bandwidth at 49GB/s and 700MB/s. The vertical dashed lines indicate when intermediate results do not fit in DRAM for binary join. The horizontal dashed line indicates speedup of 1.
  (f) Speedup of Self linear 3-way join vs. cascaded binary join with different DRAM bandwidth. 
  (g) Performance of Star 3-way join with varying $d$ and
  $h_{bkt}$. 
  (h,i) Speedup of 3-way join vs.cascaded binary joins with $d$ and $K$ at different DRAM bandwidth. 
  }
\label{fig:perf}
\end{figure}

Fig~\ref{fig:perf} (c) shows the speedup of binary self join on the accelerator over CPU with varying sizes of the relations and distinct values in joining columns ($d$).
Since the cascaded join is not parallelized
on the PCU, the second join of the cascaded join is compute-bound on CPU. 
On the accelerator, the total amount of parallelism is the product of the number of PCUs
with SIMD computation (a vector of size 16) within each PCU, which is $64\times 16=1024$. This shifts the performance bottleneck to DRAM for streaming in the intermediate relation.
Fig~\ref{fig:perf} (c) shows that smaller percentage of unique values, $d\%$ are associated with increasing speedup  (up to 600x) due to the large sized intermediate relation in the cacaded binary join.

\subsection{Performance Analysis of Linear Self Join}
We consider $R(AB) \bowtie S(BC) \bowtie T(CD)$, where R,S,T are copies of 
the friend-friend relations with $N$ records and $d$ distinct users (column values). 
\subsubsection{Hyper-parameter Selection}\label{sec:selfjoin3}
We shall discuss the evaluation of hyperparameter selection of algorithm described in Section~\ref{sec:linear}. Fig~\ref{fig:perf} (d)
plots the execution time of 3-way join varying with $H_{bkt}$ and $g_{bkt}$ ($h_{bkt}$ = number of PMUs).
It shows that the algorithm achieve higher speedup for larger size partition of $R$ partition (i.e. small $H_{bkt}$) while exploiting DRAM prefetching. For small $g_{bkt}$, the algorithm is compute-bound for joining buckets from three relations within PMUs (3-level nested loop). As $g_{bkt}$ increases, the compute complexity reduces with smaller of size $T$ buckets and the performance bottleneck shifts to DRAM bandwidth for streaming in $T$ records. For large values of $g_{bkt}$, the $S_{ij}$ bucket within each PMU becomes very small (i.e. $\frac{|S|}{Hhg}$), resulting in very poor DRAM performance for loading $S_{ij}$. Although some PCU might
have empty $S_{ij}$ bucket, the algorithm has to wait for completion from other PCUs with non-empty $S_{ij}$ buckets because all PCUs shares the streamed $T$ records.
This synchronization and poor DRAM performance on $S_{ij}$ bucket eventually increases execution time dramatically when $g_{bkt}$ becomes too large.
\subsubsection{3-way Join vs. Cascaded Binary Joins}
Fig~\ref{fig:perf} (e) and (f) shows the speedup of 3-way join over cascaded binary joins with varying average friends per person
($f = \frac{N}{d}$), and DRAM bandwidth on the accelerator. When relation size (N) is small, 3-way join achieves up to 15x performance
advantage over binary-join because the latter is heavily IO-bound compared to compute-bound 3-way join, and the accelerator favors compute bound operations.
However, the speedup decreases with increase in relation size, $N$. Because the compute complexity of 3-way join increases quadratically with $N$, whereas, size of intermediate relation of the cascaded binary joins increases quadratically with $N$. When the intermediate relation fails to fits in DRAM, the off-chip bandwidth
drops from 49GB/s to 700MB/s, which is shown as a step increase in the speedup of 3-way over the binary join in ~\ref{fig:perf} (e) and (f). With
more friends per person, the performance cliff happens at smaller relation size. (f) shows that the advantage of
3-way join is more significant when intermediate result fit as binary-join will be more DRAM-bandwidth bounded for smaller DRAM; and less significant when the intermediate result does not fit, at which point, binary-join
will be SSD bandwidth-bounded, whereas 3-way join can still benefit from higher DRAM bandwidth.

\subsection{Performance Analysis of Linear Star 3-way Join}
Now we consider a special case of linear join where $R$ and $T$ relations are small enough to fit on-chip\footnote{With plasticine,
this means the dimensions relations are on the order of millions of records.}. Now we only need one level
of hash functions on both columns $B$ and $C$, naming $h(B)$ and $g(C)$. The only difference between cascaded binary joins and
3-way join is that binary join only performs one hash function at a time, which allow $h=g=U$. For 3-way join, we map a $(h(b),g(c))$ hash value pair to each PMU, which restricts number of buckets to 
$hg=U$. For both 3-way and cascaded binary joins, we first load $R$ and $T$ on-chip, compute hash functions on the fly, and distribute the records to PMUs with corresponding assigned hash values (in binary join) or hash value pairs (in 3-way join).
Next, we stream $S$, compute hash values and distribute to the corresponding PMUs, where the inner join
is performed.

Fig~\ref{fig:perf} (g) shows the execution time of the 3-way join with varying $h_{bkt}$ (Note, $h_{bkt}$ must be
dividable by $U$ to achieve the maximum $hg$). Fig~\ref{fig:perf} (h) and (i) shows the speedup of 3-way
join over a cascade of binary star join. We can see that with increasing DRAM-bandwidth, the advantage of 3-way join eventually
disappears since storing and loading intermediate results in binary join becomes free, when they fit on the chip. 3-way join can also be slower than binary join for larger 
number of buckets (ie. less computation), where number of buckets is $hg=U^2$ for binary and $hg=U$ for 3-way join join\footnote{Total amount of
comparison in cascaded binary join roughly equals to $\frac{|R||S|}{h} + \frac{|R\bowtie S||T|}{g} = \frac{|R||S|}{h} + \frac{|R||S||T|}{dg}$}).

%% file: conclusions.tex
\section{Conclusions} \label{sec:conclusions}
Multiway join involves joining of multiple relations simultaneously instead of traditional cascaded binary joins of relations. In this paper, we have presented algorithms for efficient implementation of linear and cyclic multiway joins using coarse grain configurable accelerator such as Plasticine, which is designed for compute-intensive applications and high on-chip network communication. 
The algorithms have been discussed with their cost analysis in the context of three relations (i.e. 3-way join).

The performance of linear 3-way joins algorithms are compared to the cascaded binary joins using performance model of the Plasticine-like accelerator.
We have shown 200X to 600X improvements for traditional cascaded binary joins on the accelerator over CPU systems.
We have concluded that 3-way join can provide higher speedup over cascaded binary joins in a DRAM bandwidth-limited system or with relations having low distinct column values ($d$) (which results in
large size intermediate relation). In fact,
the effective off-chip bandwidth will dramatically reduce when the intermediate size does not fit in DRAM, in which case binary join will provide a substantial improvement over 3-way join.
We have shown that a Self 3-way join (e.g, friends of friend query) is 45X better than a traditional two cascaded binary joins for as large as 200 million records with 700 thousand distinct users. A  data-warehouse Star 3-way join query is shown to have 11X better than that of cascaded binary joins. 

In future work, we would like to explore additional levels of hashing beyond two levels, and
exploring new algorithms, such as set value join~\cite{emptyheaded}, within on-chip join to 
speedup multi-way join.
We plan to extend the algorithms for skewed data distribution in relations and analyze the
improvements in the performance and power of the algorithms on Plasticine accelerator.

%% file: model.tex
\section{Performance Model of Plasticine} \label{sec:model}

\begin{figure}
  \includegraphics[width=1\columnwidth, trim={0 0.5in 0 0}, clip]{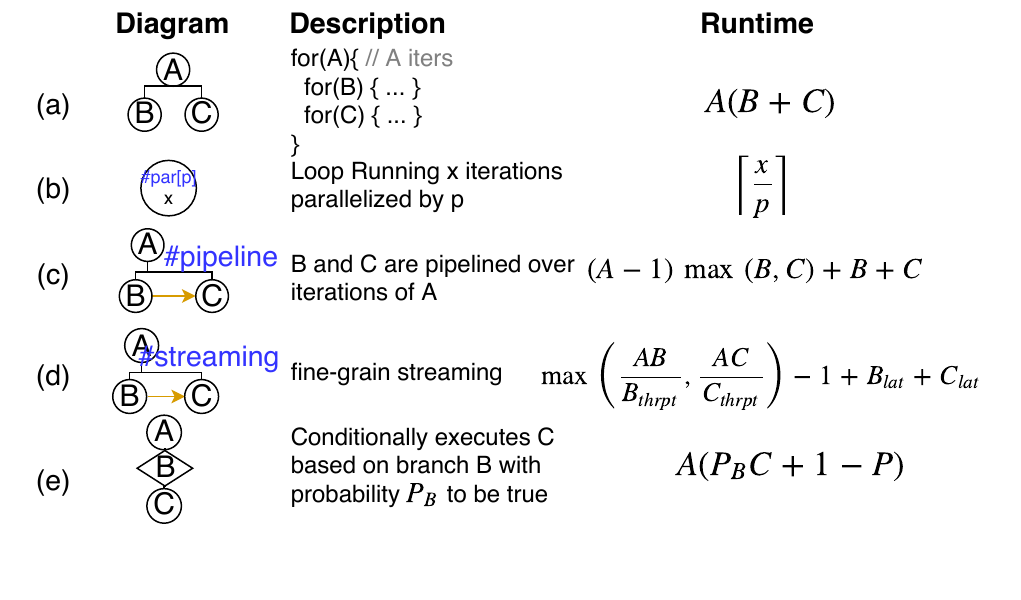}
\caption{Runtime model for different loop schedule.}
\label{fig:runtime}
\includegraphics[width=1\columnwidth]{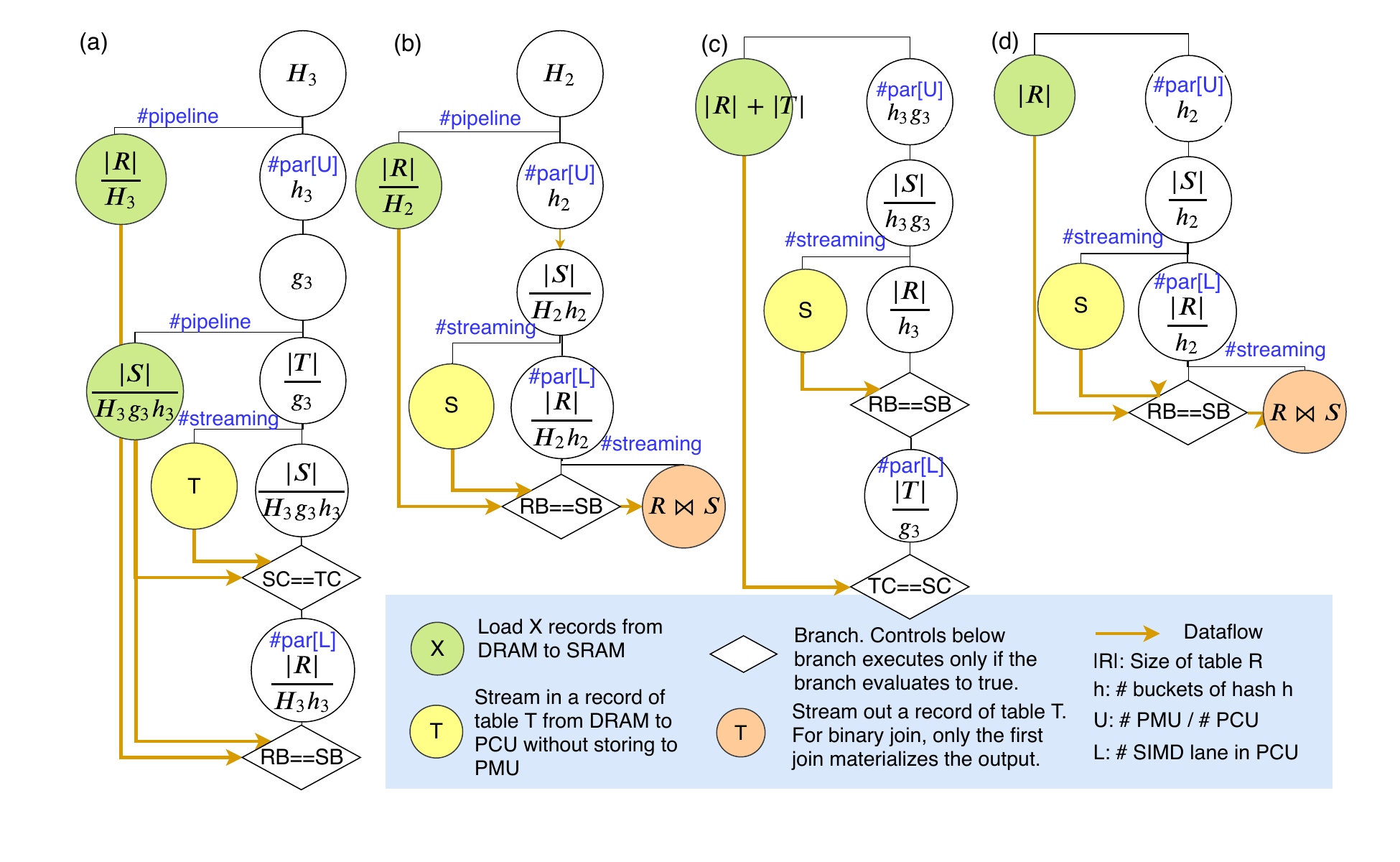}
  \caption{Loop structure of (a,b) 3-way and cascaded binary self join and (c,d)
  3-way and cascaded binary star join. Data reorganization is not shown. Only one of the join 
  in cascaded binary join is shown in (b) and (d). }
\label{fig:loop}
\vspace{-20pt}
\end{figure}

In this section, we provide more details on the analytical performance model used for algorithm performance estimation on Plasticine-like accelerator.
The performance model analyzes the loop structures of
each algorithm, takes into account how applications are spatially parallelized
and pipelined on hardware resource, and provides a cycle-level runtime estimation
given data characteristics and architectural parameters as inputs.
Fig.~\ref{fig:loop} shows the loop structures of 3-way and cascaded binary self and star join 
algorithms on the accelerator.
To avoid confusion, we use $\langle hash\rangle _2$ and $\langle hash\rangle _3$ for hash functions of binary and 3-way joins- they do not need to be the same.

In Fig.~\ref{fig:runtime} (a), the circles indicate one-level of loop nest, and the hierarchy indicates the
nest levels between loops. \ttblue{\#par[P]} in ~\ref{fig:runtime} (b) suggests a loop parallelized by P.
\ttblue{\#pipeline} in Fig~\ref{fig:runtime} (c) indicates overlapping execution of the inner loops across iterations of the outer loop,
e.g. B can work on the second iteration of A while C is working on the first iteration of A.
The pipeline construct is commonly used when a tile of data is reused multiple times on-chip, 
in which we can overlap prefetching of future tiles with execution of the current tile.
In contrary, \ttblue{\#streaming} in ~\ref{fig:runtime} (d) indicates fine-grain pipelining between producer and consumer loops, where
the consumer loop only scans the data once without any reuse.
In such case, C can execute as soon
as B produces the first chunk of data, without waiting for B to finish on one entire iteration of A.

On Plasticine-like acelerator, an example of the streaming construct is streaming data from DRAM directly to PCUs without storing to PMUs.
To compute execution time (or run time) , we need the throughput (thrpt) and latency (lat) of which B and C produces/consumes data chunks.
For DRAM, throughput and latency can be derived from DRAM bandwidth and response time, respectively.
For loops executed on Plasticine, throughput is the amount of allocated parallelism between ($U$) and within PCUs ($L$).
We used $U=64$ PCUs and SIMD vector width $L=16$ in our evaluation. The latency is the sum of network latency (we used
the worst diagonal latency on a $16\times 8$ chip, which is 24 cycles) and pipeline latency of the PCU (6 cycles).
The overall runtime of the outer loop is bounded by the stage with minimum throughput.

Finally, for data-dependent execution in ~\ref{fig:runtime} (d), we compute runtime by associating a probability to each branch. 
For example, in Fig.~\ref{fig:loop} (a), the branch on $SC==TC$ indicates comparisons on S records with streamed T records.
Only matches records will be compared with R records. The probability of this branch is the expected size of $S\bowtie T$,
which is $\frac{|S||T|}{d}$, 
over the total number of comparisons performed between S and T records. The number of comparison is the product of loop iterations
enclosing the branch, which is $H_3h_3g_3\frac{|T|}{g_3}\frac{|S|}{H_3g_3h_3} = \frac{|S||T|}{g_3}$. This gives the probability of $\frac{g_3}{d}$ on the branch hit.

Using a similar approach, we can derive probabilities of all data-dependent branches. 
The runtime of each algorithm in Fig.~\ref{fig:loop}
is recursively evaluated at each loop level using equations shown in Fig.~\ref{fig:runtime}.
The exact model is open-source and can be found at 
\url{https://github.com/yaqiz01/multijoin_plasticine.git}.